\documentclass[12pt,usenames,dvipsnames]{article}

\usepackage{ragged2e}
\usepackage{sectsty}

\usepackage{amsmath,amsthm,amssymb,amscd,mathtools,bm,upgreek}
\usepackage{graphicx}
\usepackage{epsfig}
\usepackage{hyperref}
\hypersetup{colorlinks,linkcolor=blue,citecolor=blue,urlcolor = blue}
\usepackage{rotating}
\usepackage{float}
\usepackage{enumitem}
\usepackage{multirow}
\usepackage{booktabs}
\usepackage{eurosym}
\usepackage{caption}
\usepackage{pdfcolfoot}
\usepackage{longtable}
\usepackage[title]{appendix}
\RequirePackage{mathtools,algpseudocode,etoolbox}
\usepackage[linesnumbered,ruled,vlined]{algorithm2e}

\usepackage{listings}
\usepackage{color} 
\usepackage[style=apa,backend=biber]{biblatex}
\DeclareSourcemap{
  \maps[datatype=bibtex]{
    \map{
      \step[fieldset=issn, null]
      \step[fieldset=isbn, null]
      \step[fieldset=doi, null]
      \step[fieldset=url, null]
      \step[fieldset=note, null]
    }
  }
}
\usepackage{subcaption}
\usepackage{mwe}

\usepackage{siunitx}

\usepackage{xcolor,soul}
\definecolor{blueb}{RGB}{79,112,190}

\usepackage{authblk}

\def\vx{{\boldsymbol{x}}}

\def\comp{\raise 1pt \hbox{$\scriptstyle\circ$}}

\def\upto{{\raise 1pt \hbox{$\scriptstyle \,\nearrow\,$}}}
\def\downto{{\raise 1pt \hbox{$\scriptstyle \,\searrow\,$}}}

\newtheorem{definition}{Definition}

\title{\Large\textbf{Piece-wise linear isotonic regression}
}

\author[1]{Timo Kuosmanen}
\author[2]{Juan F. Monge}
\author[2]{José L. Ruiz}
\author[3,\footnote{
Corresponding author. \newline \hspace*{5mm} 
\textit{E-mail addresses:} \texttt{timo.kuosmanen@utu.fi (T. Kuosmanen)}, \texttt{monge@umh.es (J.F. Monge)},\\
    \hspace*{34mm} \texttt{jlruiz@umh.es (J.L. Ruiz)}, \texttt{x.zhou@surrey.ac.uk (X. Zhou)}.}]{Xun Zhou}
\affil[1~]{Turku School of Economics, University of Turku, 20500 Turku, Finland}
\affil[2~]{Miguel Hernández University of Elche, 03202 Elche, Alicante, Spain}
\affil[3~]{{Surrey Business School, University of Surrey, Guildford GU2 7XH, UK}}
\date{\today}

\begin{document}

\maketitle

\vfill
\begin{center}
Declarations of interest: none
\end{center}
\vfill

\begin{abstract}
\noindent 
Isotonic regression provides a flexible, tuning-free approach to estimating monotonic functions without imposing global curvature constraints, yet the estimated regression function is inherently a step function. This paper addresses a key limitation of such estimators: their inability to provide meaningful marginal properties, such as shadow prices or elasticities. We propose a novel piece-wise linear smoothing framework that recovers meaningful marginal estimates even in non-convex settings. Building on the concept of conditional convexity originally developed in deterministic frontier analysis, we formulate the smoothing process as a bilevel optimization problem that fits a continuous, monotonic, piece-wise linear function to the initial isotonic regression predictions. Monte Carlo simulations demonstrate that the proposed approach can significantly improve estimation accuracy in both convex and non-convex settings for univariate and multivariate data. We apply this approach to analyze agglomeration economies in Finnish municipalities, illustrating its practical value.
\\[5mm]
\end{abstract}
\vfill

\thispagestyle{empty}
\newpage
\setcounter{page}{1}
\setcounter{footnote}{0}
\pagenumbering{arabic}
\baselineskip 20pt
\setlength\bibitemsep{1.15\itemsep}

\section{Introduction}\label{sec:intro}
Tuning-free, shape-constrained nonparametric regression has a long history, with foundational contributions dating back to the seminal works of \textcite{hildreth_point_1954} and \textcite{brunk_maximum_1955}. In recent years, this approach has gained renewed interest across economics \parencite{dai_can_2025,fang_projection_2021}, statistics \parencite{feng_nonparametric_2022,mukherjee_least_2024}, and operational research \parencite{Kuosmanen2021,curmei_shape-constrained_2025}. Following the piece-wise linear characterization of the multivariate convex nonparametric least squares (CNLS) estimator by \textcite{Kuosmanen2008}, much of this research has focused on the estimation of concave, monotonic increasing regression functions \parencite[e.g.,][]{dai_optimal_2025, rodseth_combining_2025,luo_sparse_2026,wang_nonparametric_2026}.

However, the assumption of global concavity is often too restrictive for empirical applications where the underlying production or utility function may exhibit non-convexities (e.g., S-shaped curves). To relax these global curvature constraints, \textcite{keshvari_stochastic_2013} develop an isotonic nonparametric least squares (INLS) estimator, which fits a monotonic increasing step function to the data. While INLS effectively ensures monotonicity without imposing concavity, the resulting step function presents a significant limitation: it is not differentiable. In many economic and operational applications, one is not only interested in the point estimates of the function itself but also in its marginal properties, such as shadow prices or elasticities, which cannot be evaluated using a step function. Furthermore, applying standard convex interpolation to smooth these steps would counterproductively reimpose the very global curvature constraints that INLS was designed to avoid, thereby erasing any identified local non-convexities.

To bridge this gap, we propose a novel piece-wise linear smoothing framework that enables the recovery of meaningful marginal properties even in non-convex settings. Building on the concept of conditional convexity, which was originally suggested by \textcite{kuosmanen_dea_2001} in the deterministic frontier setting of data envelopment analysis (DEA), we operationalize this smoothing process through a bilevel programming formulation \parencite{monge_setting_2023,monge_measuring_2025}. By doing so, our approach successfully constructs a continuous, piece-wise linear regression surface that not only strictly preserves the local shape variations and non-convexities identified by the initial INLS step function, but also directly unlocks the ability to evaluate shadow prices and elasticities.

Performance comparisons in the controlled environment of Monte Carlo simulations suggest that our piece-wise linear smoothing technique can improve estimation accuracy. This improvement is observed in both convex and non-convex cases, as well as across univariate and multivariate settings. Furthermore, we demonstrate the practical value of the proposed approach with an empirical application to Finnish municipalities, examining the relationship between population size and job availability to estimate agglomeration economies.

The rest of this paper is organized as follows. Section \ref{sec:isotonic} briefly reviews the standard isotonic regression model. Section \ref{sec:piece-wise} introduces our proposed piece-wise linear isotonic regression estimator and the bilevel programming formulation. Section \ref{sec:mc-sim} presents the design and results of the Monte Carlo simulations. Section \ref{sec:appl} discusses the empirical application to Finnish municipalities. Section \ref{sec:conc} concludes with suggestions for future research. Additionally, supplementary tables are provided in the Appendix.

\section{Isotonic regression}\label{sec:isotonic}

Consider the standard nonparametric regression model:
\begin{equation}
	y_i = f(\vx_i) + \varepsilon_i,
	\label{eq:eq1}
\end{equation}
where $y_i \in \mathbb{R}$ is the dependent variable, $\vx_i \in \mathbb{R}^s$ is a vector of explanatory variables, and $\varepsilon_i$ is a random error term with $E(\varepsilon_i) = 0$ and $Var(\varepsilon_i) = \sigma^2$. The key assumption in isotonic regression is that the function $f: \mathbb{R}^s \to \mathbb{R}$ is a monotonic increasing (or decreasing) regression function with an unknown functional form \parencite{keshvari_stochastic_2013}.

Formally, this monotonicity is grounded in the concept of isotonicity. A function $f$ is defined as isotonic with respect to a partial order $\preccurlyeq$ on the domain $\mathbb{R}^s$ if $\vx_i \preccurlyeq \vx_j$ implies $f(\vx_i) \le f(\vx_j)$. In production analysis and economics, it is typically specified as the standard dominance relation (i.e., $\vx_i \le \vx_j$ component-wise), which corresponds to the axiom of free disposability. Under this specification, monotonicity becomes a special case of isotonicity \parencite{keshvari_stochastic_2013}.

To estimate isotonic regression functions, the INLS estimator seeks to find a set of fitted values that minimize the sum of squared residuals subject to the constraints of monotonicity \parencite{keshvari_stochastic_2013}. To operationalize these constraints in a multivariate setting, a binary adjacency matrix $\mathcal{P} = [p_{ij}]_{n \times n}$ is defined such that:
\begin{equation}
    p_{ij} = 
    \begin{cases}
    1 & \text{if } \vx_i \le \vx_j, \\
    0 & \text{otherwise}.
    \end{cases}
\end{equation}

The estimation problem is formulated as the following quadratic programming (QP) problem:
\begin{align} 
    \min_{\alpha, \varepsilon} \quad 
    & \sum_{i=1}^n \varepsilon_i^2 \label{eq:inls} \\\nonumber \text{s.t.} \quad 
    & y_i = \alpha_i + \varepsilon_i, \quad \forall i = 1, \dots, n, \\\nonumber 
    & p_{ij}\alpha_i \le p_{ij}\alpha_j, \quad \forall i, j = 1, \dots, n. \nonumber 
\end{align}
The optimal fitted values $\alpha^\text{INLS}_i$ obtained from solving problem \eqref{eq:inls} form a monotonic step function. This implies that the fitted values are typically constant over blocks of observations, resulting in a regression surface that is not differentiable. Specifically, the gradient is zero within the constant blocks and undefined at the jumps. While the INLS estimator is robust and consistent, the lack of differentiability prevents the evaluation of marginal properties such as shadow prices or elasticities. This limitation serves as the primary motivation for the piece-wise linear smoothing technique developed in the next section.

\section{Piece-wise linear isotonic regression}\label{sec:piece-wise}

To address the step-function nature of the INLS estimator that precludes the estimation of marginal effects, we propose a smoothing procedure that fits a continuous, piece-wise linear function to the INLS estimates. The proposed approach is based on the notion of conditional convexity, which was originally suggested by \textcite{kuosmanen_dea_2001} in the context of DEA. To implement conditional convexity in the present context of isotonic regression, we draw upon the recent advances in bilevel programming for DEA by \textcite{monge_setting_2023,monge_measuring_2025}.

\subsection{Anchor points}\label{subsec:anchor}

The smoothing procedure begins by identifying a set of discrete points that characterize the INLS solution. Recall that the INLS estimator yields a step function where the fitted values $\alpha_i^\text{INLS}$ are constant over specific blocks of observations. We refer to the representative points of these blocks as anchor points.

To formalize the selection, let $\mathcal{K} = \{\tilde{\alpha}_1, \dots, \tilde{\alpha}_m\}$ denote the set of unique fitted values obtained from the INLS estimation, where $m$ denotes the total number of isotonic blocks. For each unique level $\tilde{\alpha}_k \in \mathcal{K}$, we define the corresponding block of observations as $\mathcal{I}_k = \{i\mid\alpha_i^\text{INLS} = \tilde{\alpha}_k\}$. The set of anchor points is then defined as $\mathcal{A} = \{(\tilde{\vx}_k, \tilde{\alpha}_k)\}_{k=1}^m$, where $\tilde{\vx}_k \in \mathbb{R}^s$ is a representative coordinate vector for the $k$-th block. We consider two primary strategies for determining $\tilde{\vx}_k$:
\begin{enumerate}
    \item \textit{Vertices:} We define $\tilde{\vx}_k$ as a synthetic vertex composed of the component-wise minimums within each block. Formally, for each dimension $r = 1, \dots, s$, the $r$-th component of the anchor point is $\tilde{x}_{kr} = \min_{i \in \mathcal{I}_k} x_{ir}$. This synthetic point represents the lower-left boundary of the $k$-th isotonic step in the multi-dimensional input space. Using vertices focuses the interpolation on the points where each response level first becomes attainable according to the data.
    \item \textit{Centroids:} Alternatively, we calculate $\tilde{\vx}_k$ as the representative center of each block, such as the mean or median of $\vx_i$ for $i \in \mathcal{I}_k$. This strategy is designed to mitigate the potential upward bias that might arise from relying solely on the extreme lower boundaries (vertices) of the steps. 
\end{enumerate}
In the special case of a univariate regression ($s=1$), the synthetic vertex $\tilde{x}_k$ simplifies to the minimum covariate value within the block $\mathcal{I}_k$, while the centroid corresponds to the central location of the step along the single input axis.

\subsection{Conditional convexity}

With the set of anchor points $\mathcal{A}$ established, our goal is to construct a continuous regression surface by interpolating between them. A common approach in nonparametric estimation is to compute the convex hull of the data points, which results in a globally concave or convex function. However, such unrestricted convex interpolation would effectively reimpose the global curvature constraints that INLS was designed to relax, potentially flattening S-shaped patterns or other local non-convexities identified in the initial INLS estimation.

To achieve piece-wise linear smoothing while strictly preserving the local features, we adopt the notion of conditional convexity. In this context, a set of points is conditionally convex if the local linear facets used for interpolation do not dominate any existing anchor point. Intuitively, dominance implies that a virtual combination achieves a response value that is at least as high as an existing anchor, using explanatory variables that are less than or equal to those of the anchor, without being completely identical across all dimensions. This can be formally defined as follows.

\begin{definition}[Dominance]\label{def:dominance}
A subset of anchor points $\mathcal{R} \subseteq \mathcal{A}$ is said to dominate an anchor point $k \in \mathcal{A}$ if and only if there exists a convex combination $(\vx, \alpha)$ of the points in $\mathcal{R}$ such that $(\vx,-\alpha) \le (\tilde{\vx}_k,-\tilde{\alpha}_k)$ and $(\vx,-\alpha) \neq (\tilde{\vx}_k,-\tilde{\alpha}_k)$.
\end{definition}

\begin{definition}[Conditional convexity]\label{def:cc}
A subset $\mathcal{R} \subseteq \mathcal{A}$ satisfies conditional convexity if and only if $\mathcal{R}$ does not dominate any existing anchor point $\in \mathcal{A}$.
\end{definition}

While conditional convexity was originally introduced in the context of DEA (best-practice or boundary seeking) to preserve efficiency classifications in non-convex production sets, we utilize it here for the distinct purpose of shape-preserving smoothing in the context of isotonic regression (average-practice or mean approximating). Conditional convexity is adapted to ensure that the smoothing function adheres to local non-convex patterns, such as the transition around the inflection point of an S-shaped curve, rather than bridging over them. This allows the smoothing technique to recover meaningful marginal estimates (e.g., shadow prices or elasticities) that vary correctly across the regression surface \parencite{monge_measuring_2025}.

\subsection{Bilevel programming}

Having derived the set of anchor points, we proceed to construct a continuous regression estimator $\hat{y}(\vx)$ defined over the entire domain. For the explanatory variables $\vx_0 \in \mathbb{R}^s$ of any given observation, calculating the predicted response $\hat{y}(\vx_0)$ requires finding a valid local interpolation based on the anchors in $\mathcal{A}$. To implement conditional convexity, we must select a reference subset of anchor points that forms a local linear facet enclosing $\vx_0$ without dominating any other existing anchors in $\mathcal{A}$. This selection process is automated by solving a bilevel optimization problem for each $\vx_0$ \parencite{monge_setting_2023,monge_measuring_2025}. 

Let $\theta_0$ represent the predicted response value at $\vx_0$. We introduce binary variables $b_j \in \{0, 1\}$ for each anchor point $j \in \mathcal{A}$, where $b_j = 0$ indicates that anchor $j$ is active (selected to form the interpolation facet). Adapting from \textcite{monge_setting_2023,monge_measuring_2025}, the conditionally convex INLS (CC-INLS) estimator $\hat{y}(\vx_0) = \theta_0^*$ is obtained by solving:
\begin{align}
    \max_{\theta, \lambda, b} \quad
    & \theta_0 + \delta \sum_{j \in \mathcal{A}} b_j \label{eq:leader} \\\text{s.t.} \quad
    & \sum_{j \in \mathcal{A}} \lambda_j \tilde{\vx}_j \le \vx_0, \label{eq:leader_x} \tag{4.1} \\
    & \sum_{j \in \mathcal{A}} \lambda_j \tilde{\alpha}_j \ge \theta_0, \label{eq:leader_y} \tag{4.2}\\
    & \sum_{j \in \mathcal{A}} \lambda_j = 1, \label{eq:leader_sum} \tag{4.3}\\
    & \lambda_j \le 1 - b_j, \quad \forall j \in \mathcal{A}, \label{eq:leader_bin} \tag{4.4}\\
    & -\boldsymbol{v}^\top \tilde{\vx}_j + u \tilde{\alpha}_j + u_0 + d_j = 0, \quad \forall j \in \mathcal{A}, \label{eq:hyperplane_def} \tag{4.5}\\
    & \lambda_j d_j = 0, \quad \forall j \in \mathcal{A}, \label{eq:coplanar_condition} \tag{4.6}\\
    & \tau_k \le 0, \quad \forall k \in \mathcal{A}, \label{eq:leader_cc_constraint} \tag{4.7}\\
    & \lambda_j \ge 0, \quad b_j \in \{0, 1\}, \quad d_j \text{ free}, \quad \forall j \in \mathcal{A}, \label{eq:leader_domain}\tag{4.8}\\
    & \theta_0, u_0 \text{ free}, \quad \boldsymbol{v} \ge \mathbf{1}_s, \quad u \ge 1, \label{eq:leader_domain2}\tag{4.9}
\end{align}
where $\tau_k$ refers to the optimal objective values of the follower problem:
\begin{align}
    \max_{\tau, \lambda, t} \quad
    & \sum_{k \in \mathcal{A}} \tau_k \label{eq:follower} \\\text{s.t.} \quad
    & \sum_{j \in \mathcal{A}} \lambda_j^k \tilde{\vx}_j = \tilde{\vx}_k - \boldsymbol{t}_k^{-}, \quad \forall k \in \mathcal{A}, \label{eq:follower_x} \tag{5.1}\\
    & \sum_{j \in \mathcal{A}} \lambda_j^k \tilde{\alpha}_j = \tilde{\alpha}_k + t_k^{+}, \quad \forall k \in \mathcal{A}, \label{eq:follower_y} \tag{5.2}\\
    & \sum_{j \in \mathcal{A}} \lambda_j^k = 1, \quad \forall k \in \mathcal{A}, \label{eq:follower_sum} \tag{5.3}\\
    & \lambda_j^k \le 1 - b_j, \quad \forall j, k \in \mathcal{A}, \label{eq:follower_bin} \tag{5.4}\\
    & \boldsymbol{t}_k^{-} \ge \tau_k\mathbf{1}_s, \quad t_k^{+} \ge \tau_k, \quad \forall k \in \mathcal{A}, \label{eq:follower_tau} \tag{5.5}\\
    & \lambda_j^k \ge 0, \quad \boldsymbol{t}_k^{-}, t_k^{+}, \tau_k \text{ free}, \quad \forall j, k \in \mathcal{A}.\label{eq:follower_domain} \tag{5.6}
\end{align}

This bilevel problem \eqref{eq:leader}--\eqref{eq:follower} models a game between a leader (who proposes a prediction) and a follower (who tests if the prediction satisfies conditional convexity). The leader problem \eqref{eq:leader} constructs the prediction $\theta_0$ for $\vx_0$. Its objective function prioritizes maximizing $\theta_0$ to find the closest possible fit on the regression surface. Constraints \eqref{eq:leader_x}--\eqref{eq:leader_sum} ensure that the evaluation point $(\vx_0,\theta_0)$ is a convex combination of anchor points, namely those with $\lambda_j > 0$ (the active anchors). Constraints \eqref{eq:hyperplane_def}--\eqref{eq:coplanar_condition} constitute the core mechanism for ensuring that the selected active anchors are coplanar. Specifically, constraint \eqref{eq:hyperplane_def} defines a global supporting hyperplane parameterized by the input multiplier vector $\boldsymbol{v}$, the output multiplier $u$, and the intercept $u_0$, where $d_j$ represents the slack distance from anchor $j$ to this hyperplane. Constraint \eqref{eq:coplanar_condition} introduces an exact, nonlinear complementary slackness condition to enforce coplanarity. If an anchor is assigned a strictly positive weight ($\lambda_j > 0$), this constraint strictly forces $d_j = 0$, binding the active anchor to the supporting hyperplane. Conversely, if an anchor is inactive ($\lambda_j = 0$), it is permitted to reside freely on either side of the local supporting facet, which is essential for penetrating through local non-convexities rather than strictly enveloping them.

The follower problem \eqref{eq:follower} functions as a rigorous conditional convexity test. The binary variables $b_j$'s both in the leader and in the follower are used as indicators for identifying the active anchors. Note that in maximizing them through the secondary term of the objective function of the leader, $\delta \sum_{j \in \mathcal{A}} b_j$, where $\delta$ is a non-Archimedean infinitesimal, we have that an anchor is active if, and only if, its $b_j=0$ (and, consequently, the intensity weights $\lambda_j^k$ are zero for any anchor not selected). Therefore, we can be sure that both problems, the leader and the follower, consider convex combinations of the same set of active anchors at optimum, which is crucial for modeling. For the specific facet proposed by the leader, the follower attempts to invalidate the selection by constructing a virtual point, a convex combination of the active anchors, that dominates any existing anchor $k \in \mathcal{A}$. Analogous to the leader problem, constraints \eqref{eq:follower_sum}--\eqref{eq:follower_bin} ensure that this virtual point is a valid convex combination derived solely from the active anchors. Constraints \eqref{eq:follower_x}--\eqref{eq:follower_y} define the slack variables $\boldsymbol{t}_k^- \in \mathbb{R}^s$ and $t_k^+$, which represent potential contraction of $\tilde{\vx}_k$ and expansion of $\tilde{\alpha}_j$, respectively. By maximizing the dominance indicator $\tau_k$, defined in \eqref{eq:follower_tau} as the minimum of these slacks, the follower actively seeks evidence that the proposed facet passes above a real data point $k$. A strictly positive objective value ($\tau_k^* > 0$) would imply that the virtual point dominates the real anchor, indicating that the interpolation surface bridges a non-convex region, such as the inflection point of an S-curve. Conversely, a strictly negative value ($\tau_k^* < 0$) guarantees that the virtual point is inferior to the anchor in at least one dimension, confirming that no dominance occurs and hence conditional convexity is met.

However, a critical boundary case arises if the follower problem yields $\tau_k^* = 0$. In this scenario, evaluating conditional convexity requires a strict inspection of the slacks ($\boldsymbol{t}_k^-$ and $t_k^+$). If all slacks are exactly zero, the virtual point simply coincides with the existing anchor. But if $\tau_k^* = 0$ alongside any strictly positive slack, the virtual point achieves a strict improvement in at least one dimension without being inferior in any other. This suggests that the virtual point still dominates the anchor, thereby violating conditional convexity \parencite{monge_measuring_2025}.

Crucially, the interaction between the two levels is governed by constraint \eqref{eq:leader_cc_constraint}, $\tau_k \le 0$, in the leader problem. This constraint prevents the leader from selecting any set of anchors that would allow the follower to find a positive $\tau_k$, thereby enforcing conditional convexity. It does, however, mathematically permit the boundary case of $\tau_k = 0$. If this boundary case were to occur, a post-estimation check of the slacks would be required to rule out any strictly positive slack, which would constitute a violation of conditional convexity. \textcite{monge_measuring_2025} state that this issue rarely arises in practice but, if it does, it is easy to handle: we can add constraints that prevent convex combinations of the current set of active anchors and solve the resulting model. It has been observed that we can find an optimal solution satisfying the desired requirements after solving very few times the extended model (see that paper for details).

Finally, constraints \eqref{eq:leader_domain}--\eqref{eq:leader_domain2} and \eqref{eq:follower_domain} specify the variable domains. The selection variables $b_j$ are binary, while the intensity weights $\lambda_j$ are non-negative. Notably, the multiplier constraints $\boldsymbol{v} \ge \mathbf{1}_s$ and $u \ge 1$ explicitly enforce monotonicity, ensuring that the constructed enveloping facets are upward-sloping, which is essential for capturing standard production axioms. In contrast, the intercept $u_0$, the slack distances $d_j$, and the predicted response $\theta_0$, as well as the follower's slack variables $\boldsymbol{t}_k^-, t_k^+$ and dominance indicator $\tau_k$, are free in sign.

\subsection{Single-level reformulation}

The bilevel programming problem \eqref{eq:leader}--\eqref{eq:follower} is computationally challenging to solve directly. To implement the CC-INLS estimator practically, we reformulate it into a single-level mixed-integer linear programming (MILP) problem, adapting the framework of \textcite{monge_measuring_2025}. 

The reformulation relies on two key transformations. First, we need to linearize the exact coplanarity condition, $\lambda_j d_j = 0$, in the leader problem \eqref{eq:leader}. This can be achieved by leveraging the existing binary selection variables $b_j$ and the bounding constraint $\lambda_j \le 1 - b_j$. Since $\lambda_j > 0$ strictly implies $b_j = 0$, the complementary slackness condition is logically equivalent to enforcing $d_j = 0$ whenever $b_j = 0$. Because $d_j$ is unrestricted in sign, this conditional logic translates perfectly into the two-sided Big-M constraints: $-M b_j \le d_j \le M b_j$, where $M$ is a sufficiently large positive constant. If $b_j = 0$ (the anchor is active), it strictly forces $d_j = 0$; if $b_j = 1$ (the anchor is inactive), $d_j$ is completely free to take any necessary value.

Second, we collapse the bilevel architecture by replacing the follower problem \eqref{eq:follower} with the corresponding optimality conditions from the strong duality in linear programming (LP). To handle the coupling constraint \eqref{eq:follower_bin}, $\lambda_j^k \le 1 - b_j$, we move it into the follower's objective function using a Big-M penalty. The exact penalized objective function that corresponds to problem \eqref{eq:follower} becomes:
\begin{equation}
    \max_{\tau, \lambda, t} \quad \sum_{k \in \mathcal{A}} \left( \tau_k - M \sum_{j \in \mathcal{A}} b_j \lambda_j^k \right),
    \label{eq:follower_primal_penalized}
\end{equation}
subject to the same constraints as in problem \eqref{eq:follower}, except for \eqref{eq:follower_bin}. When the leader deselects an anchor ($b_j = 1$), any assignment of a strictly positive weight ($\lambda_j^k > 0$) by the follower would trigger an overwhelming reduction in its maximization objective due to the sufficiently large constant $M$; hence, the optimal response of the rational follower is strictly forced to yield $\lambda_j^k = 0$. When the anchor is selected ($b_j = 0$), the penalty identically vanishes. 

Because the optimization problems involved for the different $k$'s are completely separable, for any fixed binary selection $b_j$ made by the leader, analyzing the global problem is equivalent to analyzing the individual sub-problems for each $k$. Therefore, for a given existing anchor $k \in \mathcal{A}$, the penalized follower problem reduces to a pure LP problem with the following objective:
\begin{equation}
    \max_{\tau, \lambda, t} \quad \tau_k - M \sum_{j \in \mathcal{A}} b_j \lambda_j^k.
    \label{eq:follower_primal_penalized_2}
\end{equation}

Let $\boldsymbol{v}_k \in \mathbb{R}^s$, $u_k \in \mathbb{R}$, and $w_k \in \mathbb{R}$ be the unrestricted dual multipliers associated with the input, output, and convexity equality constraints \eqref{eq:follower_x}--\eqref{eq:follower_sum}, respectively. Furthermore, let $\boldsymbol{\nu}_k \ge \mathbf{0}$ and $\mu_k \ge 0$ be the non-negative dual multipliers corresponding to \eqref{eq:follower_tau}, which are naturally rewritten in the standard form as $-\boldsymbol{t}_k^- + \tau_k \mathbf{1}_s \le \mathbf{0}$ and $-t_k^+ + \tau_k \le 0$. The Lagrangian $L$ of problem \eqref{eq:follower_primal_penalized_2} is formulated as follows:
\begin{align}
L(\lambda, \boldsymbol{t}_k^-, t_k^+, \tau_k, \boldsymbol{v}_k, u_k, w_k, \boldsymbol{\nu}_k, \mu_k) & = (\tau_k - M \sum_{j \in \mathcal{A}} b_j \lambda_j^k) + \boldsymbol{v}_k^\top (\tilde{\vx}_k - \sum_{j \in \mathcal{A}} \lambda_j^k \tilde{\vx}_j - \boldsymbol{t}_k^-)\\\nonumber
& + u_k (\tilde{\alpha}_k - \sum_{j \in \mathcal{A}} \lambda_j^k \tilde{\alpha}_j + t_k^+) + w_k (1 - \sum_{j \in \mathcal{A}} \lambda_j^k) \\\nonumber
& + \boldsymbol{\nu}_k^\top (\boldsymbol{t}_k^- - \tau_k \mathbf{1}_s) + \mu_k (t_k^+ - \tau_k).
\end{align}

According to the standard duality theory for LP, because the primal variables $\boldsymbol{t}_k^-$, $t_k^+$, and $\tau_k$ are completely unrestricted in sign, their corresponding partial derivatives in the Lagrangian must strictly equal zero to maintain a bounded dual objective. Extracting the partial derivatives for these free variables yields the following necessary conditions:
\begin{align}
\dfrac{\partial L}{\partial \boldsymbol{t}_k^-} = \mathbf{0} \quad & \implies \quad -\boldsymbol{v}_k + \boldsymbol{\nu}_k = \mathbf{0} \quad \implies \quad \boldsymbol{v}_k = \boldsymbol{\nu}_k \ge \mathbf{0}, \label{eq:partial_t-}\\
\dfrac{\partial L}{\partial t_k^+} = 0 \quad & \implies \quad u_k + \mu_k = 0 \quad \implies \quad u_k = -\mu_k \le 0, \label{eq:partial_t+}\\
\dfrac{\partial L}{\partial \tau_k} = 0 \quad & \implies \quad 1 - \boldsymbol{\nu}_k^\top \mathbf{1}_s - \mu_k = 0 \quad \implies \quad \boldsymbol{\nu}_k^\top \mathbf{1}_s + \mu_k = 1.\label{eq:partial_tau}
\end{align}

Substituting \eqref{eq:partial_t-}--\eqref{eq:partial_t+} into \eqref{eq:partial_tau} perfectly bridges the multipliers to produce the joint normalization constraint: $\boldsymbol{v}_k^\top \mathbf{1}_s - u_k = 1$. By evaluating the Lagrangian with respect to the continuous weights $\lambda_j^k \ge 0$, we obtain the standard dual constraint: $-\boldsymbol{v}_k^\top \tilde{\vx}_j - u_k \tilde{\alpha}_j - w_k \le M b_j$. Incorporating the joint normalization constraint and the multiplier domains leads to the dual of problem \eqref{eq:follower_primal_penalized_2}:
\begin{align}
\min_{\boldsymbol{v}_k, u_k, w_k} \quad & \boldsymbol{v}_k^\top \tilde{\vx}_k + u_k \tilde{\alpha}_k + w_k \\
\text{s.t.} \quad & -\boldsymbol{v}_k^\top \tilde{\vx}_j - u_k \tilde{\alpha}_j - w_k \le M b_j, \quad \forall j \in \mathcal{A}, \label{eq:follower_dual_penalized_dual} \tag{12.1}\\
& \boldsymbol{v}_k^\top \mathbf{1}_s - u_k = 1, \label{eq:follower_dual_penalized_norm} \tag{12.2}\\
& \boldsymbol{v}_k \ge \mathbf{0}, \quad u_k \le 0, \quad w_k \text{ free}. \label{eq:follower_dual_penalized_domain} \tag{12.3}
\end{align}

To enforce conditional convexity over the entire regression surface, the optimality conditions must hold for all existing anchors. Specifically, the penalized follower reaches global optimality if and only if its primal constraints, dual constraints, and the strong duality equation (i.e., the primal objective equals the dual objective) are simultaneously satisfied. By appending these optimality conditions for all $k \in \mathcal{A}$, alongside the linearized coplanarity bounds, to the leader's primary feasible region, we effectively collapse the entire structure into a unified, single-level MILP formulation of the CC-INLS estimator:
\begin{align}
    \max \quad 
    & \theta_0 + \delta \sum_{j \in \mathcal{A}} b_j \label{eq:final_milp} \\
    \text{s.t.} \quad 
    & \eqref{eq:leader_x}-\eqref{eq:hyperplane_def}, \eqref{eq:leader_cc_constraint}-\eqref{eq:leader_domain2} \nonumber \\
    & -M b_j \le d_j \le M b_j, \quad \forall j \in \mathcal{A}, \label{eq:coplanar_condition_M} \tag{13.1}\\
    & \eqref{eq:follower_x}-\eqref{eq:follower_sum}, \eqref{eq:follower_tau}-\eqref{eq:follower_domain} \nonumber \\
    & \eqref{eq:follower_dual_penalized_dual}-\eqref{eq:follower_dual_penalized_domain}, \quad \forall k \in \mathcal{A}, \nonumber \\
    & \sum_{k \in \mathcal{A}} \left( \tau_k - M \sum_{j \in \mathcal{A}} b_j \lambda_j^k \right) = \sum_{k \in \mathcal{A}} (v_k^\top \tilde{x}_k + u_k \tilde{\alpha}_k + w_k). \tag{13.2}
\end{align}

The derivation of this single-level MILP formulation represents a key advantage of the CC-INLS framework. By explicitly leveraging the strong duality theorem to collapse the lower-level problem, we avoid the computationally expensive Karush-Kuhn-Tucker (KKT) conditions. By equating the primal and dual objectives of the follower, we obtain a unified, mathematically pure MILP that can be solved efficiently using standard solvers (e.g., CPLEX or Gurobi).

\section{Monte Carlo simulations}\label{sec:mc-sim} 

To evaluate the finite-sample performance of the proposed CC-INLS estimator, we conduct a series of Monte Carlo simulations. The primary objective is to demonstrate the estimator's capability to recover meaningful structural patterns under varying noise levels, particularly when the underlying true data-generating process (DGP) exhibits local non-convexities that traditional convex nonparametric estimators typically fail to capture.

\subsection{Simulation setup}

We replicate and adapt the experimental design originally proposed by \textcite{keshvari_stochastic_2013}. To strictly align the simulations with our regression context rather than a stochastic frontier framework, we exclude the asymmetric inefficiency term. 

Two distinct functional forms for the true regression function in equation \eqref{eq:eq1} are examined to test the flexibility of the CC-INLS estimator. The first is a logistic function, which is monotonic but non-concave, exhibiting both convex and concave regions (S-shaped). This allows us to test the estimator's robustness in settings that violate the assumptions of traditional convex regression. The second is a Cobb-Douglas function, which represents a globally concave and monotonic function characterized by diminishing marginal returns. This scenario compares how the proposed estimator performs when the underlying function satisfies standard convexity assumptions. For both functional forms, we consider specifications with a single explanatory variable ($s=1$) and two explanatory variables ($s=2$) to evaluate the performance across different dimensionalities.

Furthermore, to ensure a meaningful and comparable signal-to-noise ratio for nonparametric estimation, we specify the domain of the explanatory variables and the standard deviation of the error term ($\sigma_\varepsilon$) for each functional form. For the non-concave logistic function, the explanatory variables are drawn from a standard uniform distribution $U(0, 1)$ and we set the noise parameters as $\sigma_\varepsilon = 0.03$ and $\sigma_\varepsilon = 0.1$ for the low- and high-noise scenarios, respectively. In contrast, for the globally concave Cobb-Douglas function, the explanatory variables are drawn from a wider uniform distribution $U(1, 6)$ and we set a higher baseline noise to match its larger signal span; $\sigma_\varepsilon = 0.1$ representing the low-noise environment and $\sigma_\varepsilon = 0.2$ for the high-noise environment. 

Combining the two functional forms, two dimensionalities of explanatory variables, and two noise regimes, our experimental design encompasses a total of eight distinct simulation scenarios. Formally, the DGPs are defined as follows:
\begin{itemize}
    \item Logistic
    \begin{itemize}
        \item $s=1: y_i = \dfrac{1+e^7}{e^7}\dfrac{e^{-3+10x_i}}{1+e^{-3+10x_i}},\; x_{1i} \sim U(0, 1), \varepsilon_{i} \sim N(0, \sigma_\varepsilon^2), \sigma_\varepsilon = \{0.03,0.1\}$
        \item $s=2: y_i = \dfrac{1+e^7}{e^7}\dfrac{e^{-3+5x_{1i}+5x_{2i}}}{1+e^{-3+5x_{1i}+5x_{2i}}},\; x_{.i} \sim U(0, 1), \varepsilon_{i} \sim N(0, \sigma_\varepsilon^2), \sigma_\varepsilon = \{0.03,0.1\}$
    \end{itemize}
    \item Cobb-Douglas
    \begin{itemize}
        \item $s=1: y_i = x_{i}^{0.8} + \varepsilon_i,\; x_{i} \sim U(1, 6), \varepsilon_{i} \sim N(0, \sigma_\varepsilon^2), \sigma_\varepsilon = \{0.1,0.2\}$
        \item $s=2: y_i = x_{1i}^{0.4}x_{2i}^{0.4} + \varepsilon_i,\; x_{.i} \sim U(1, 6), \varepsilon_{i} \sim N(0, \sigma_\varepsilon^2), \sigma_\varepsilon = \{0.1,0.2\}$
\end{itemize}
\end{itemize}

Across all the simulated scenarios, the baseline estimator is the standard, unsmoothed INLS estimator defined in \eqref{eq:inls}. Against this baseline, we evaluate two variants of our proposed CC-INLS estimator applied in the smoothing stage, which differ solely in the choice of anchor points. The first variant, denoted CC-INLS (vertices), uses the vertices (i.e., the minimum covariate values) of each INLS block as anchors. The second variant, denoted CC-INLS (centroids), employs the centroids (e.g., averages) of the blocks. The finite-sample performance of each estimator is evaluated using the root mean squared error (RMSE) as the primary metric. To maintain the flow of the main text, detailed tables reporting the mean squared error (MSE) and bias are provided in Appendix \ref{sec:supplement_tables}.

\subsection{Results of the logistic scenario}

We first analyze the results for the logistic function scenario. This S-shaped function represents a challenging setting for conventional shape-constrained estimators, as estimators imposing global concavity would suffer from significant specification bias in the region of increasing returns. Table \ref{tab:rmse_logistic} summarizes the out-of-sample RMSE for the logistic DGPs, uniformly evaluated on an independent test set of size 100. 

\begin{table}[H]
  \centering
  \caption{Out-of-sample RMSE for the logistic DGPs ($n_{\text{out}}=100$).}
  \label{tab:rmse_logistic}
  \renewcommand{\arraystretch}{1.2}
  \small
  \begin{tabular}{@{}lcccccc@{}}
    \toprule
    \multirow{2}{*}{Estimator} & \multicolumn{3}{c}{Low noise ($\sigma_\varepsilon = 0.03$)} & \multicolumn{3}{c}{High noise ($\sigma_\varepsilon = 0.1$)} \\
    \cmidrule(lr){2-4} \cmidrule(lr){5-7}
    & $n_{\text{in}}=25$ & $n_{\text{in}}=50$ & $n_{\text{in}}=100$ & $n_{\text{in}}=25$ & $n_{\text{in}}=50$ & $n_{\text{in}}=100$ \\
    \midrule
    \multicolumn{7}{@{}c}{\textit{Panel A: Single explanatory variable ($s=1$)}} \\
    \midrule
    INLS  & 0.0736 & 0.0476 & 0.0374 & 0.1304 & 0.1158 & 0.1099 \\
    CC-INLS (vertices)    & 0.0481 & 0.0424 & 0.0369 & 0.1191 & 0.1132 & 0.1088 \\
    CC-INLS (centroids) & 0.0497 & 0.0445 & 0.0387 & 0.1206 & 0.1150 & 0.1107 \\
    \midrule
    \multicolumn{7}{@{}c}{\textit{Panel B: Two explanatory variables ($s=2$)}} \\
    \midrule
    INLS  & 0.2320 & 0.1814 & 0.1384 & 0.2648 & 0.2121 & 0.1832 \\
    CC-INLS (vertices)    & 0.1047 & 0.0796 & 0.0763 & 0.1642 & 0.1497 & 0.1429 \\
    CC-INLS (centroids) & 0.1070 & 0.0844 & 0.0779 & 0.1682 & 0.1494 & 0.1415 \\
    \bottomrule
  \end{tabular}
  
  \vspace{1ex}
  \footnotesize{\textit{Note:} $n_{\text{in}}$ denotes the in-sample size used for model estimation and $n_{\text{out}}$ denotes the test set size.}
\end{table}

The simulations reveal that applying conditional convexity generally enhances estimation accuracy. In almost all settings, the RMSE values for the proposed estimator (both vertices- and centroids-based) are significantly lower than those of the INLS step function. A minor exception is observed in the case of a single explanatory variable, where the improvement provided by smoothing is less pronounced. As expected, the predictive performance of all models naturally improves as the training sample size increases and degrades in higher noise environments.

Regarding the choice of anchor points, the results suggest that the specific smoothing strategy is of secondary importance. The differences in RMSE between CC-INLS (vertices) and CC-INLS (centroids) are negligible across the simulated scenarios. This indicates that the primary gain in accuracy comes from the application of the conditional convexity principle itself, i.e., restoring the continuous shape of the function, rather than the specific method used to aggregate the observations within each isotonic block.

\begin{figure}[H]
	\centering
	\begin{subfigure}[b]{0.49\textwidth}
		\centering
		\includegraphics[width=1\textwidth]{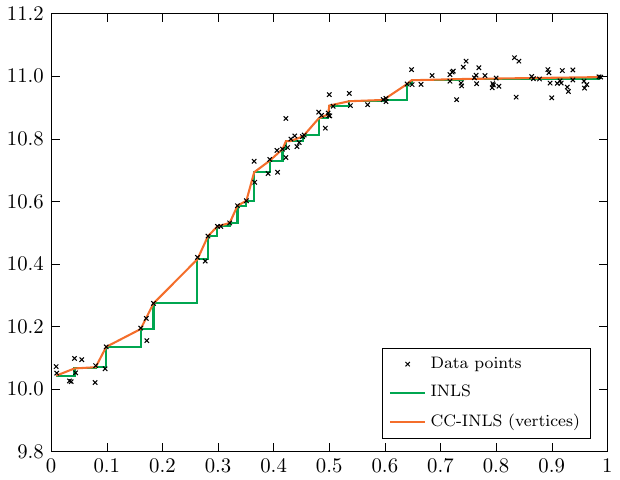} 
		\caption[]%
		{{\small Low noise (vertices)}}    
		\label{fig:log_a}
	\end{subfigure}
	\begin{subfigure}[b]{0.49\textwidth}  
		\centering 
		\includegraphics[width=1\textwidth]{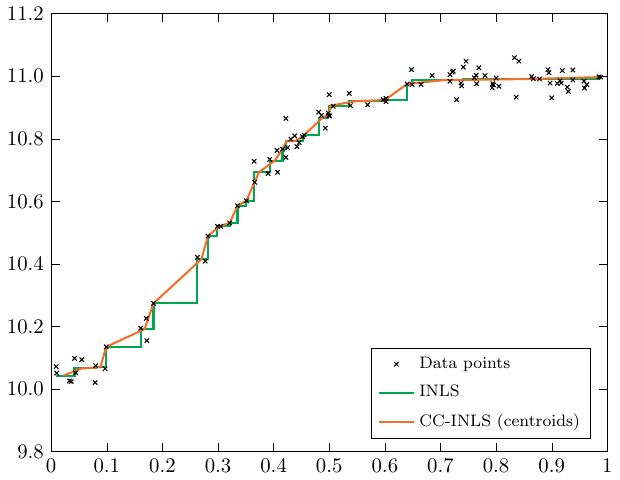} 
		\caption[]%
		{{\small Low noise (centroids)}}    
		\label{fig:log_b}
	\end{subfigure}
	\begin{subfigure}[b]{0.49\textwidth}   
		\centering 
		\includegraphics[width=1\textwidth]{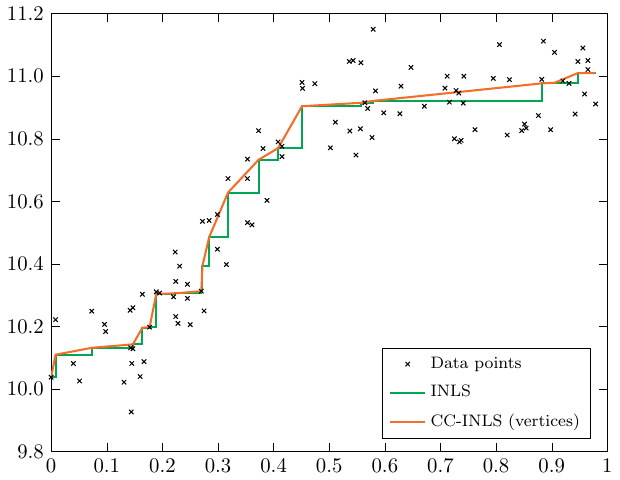} 
		\caption[]%
		{{\small High noise (vertices)}}    
		\label{fig:log_c}
	\end{subfigure}
	\begin{subfigure}[b]{0.49\textwidth}   
		\centering 
		\includegraphics[width=1\textwidth]{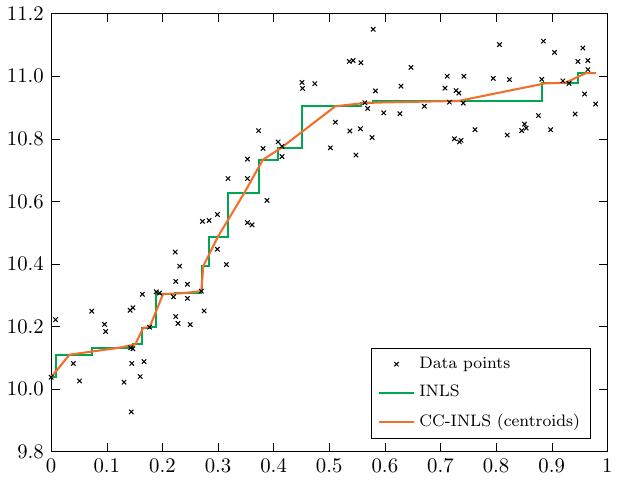} 
		\caption[]%
		{{\small High noise (centroids)}}    
		\label{fig:log_d}
	\end{subfigure}
	\caption[]%
	{\centering Illustration of the logistic functions estimated by INLS, CC-INLS (vertices), CC-INLS (centroids) across low and high noise scenarios.}
	\label{fig:log}
\end{figure}

To provide a visual comparison of the estimated regression surfaces, Figure \ref{fig:log} displays the fitted functions for a sample logistic DGP. The figure illustrates how the INLS estimator (green line) yields a step function, whereas both variants of the CC-INLS estimator (orange line) construct a continuous piece-wise surface that directly captures the underlying non-convex pattern across both low and high noise scenarios.

\subsection{Results of the Cobb-Douglas scenario}

Next, we consider the Cobb-Douglas function scenario. Table \ref{tab:rmse_cd} presents the out-of-sample RMSE for these DGPs. Consistent with the logistic scenario, both CC-INLS (vertices) and CC-INLS (centroids) systematically outperform the baseline INLS across all settings. Similarly, the choice between vertices- and centroids-based anchor points yields virtually identical predictive accuracy in this scenario. As expected, the predictive accuracy of all estimators improves with larger in-sample sizes and degrades as the noise level increases from $\sigma_\varepsilon = 0.1$ to $\sigma_\varepsilon = 0.2$.

\begin{table}[H]
  \centering
  \caption{Out-of-sample RMSE for the Cobb-Douglas DGPs ($n_{\text{out}}=100$).}
  \label{tab:rmse_cd}
  \renewcommand{\arraystretch}{1.2}
  \small
  \begin{tabular}{@{}lcccccc@{}}
    \toprule
    \multirow{2}{*}{Estimator} & \multicolumn{3}{c}{Low noise ($\sigma_\varepsilon = 0.1$)} & \multicolumn{3}{c}{High noise ($\sigma_\varepsilon = 0.2$)} \\
    \cmidrule(lr){2-4} \cmidrule(lr){5-7}
    & $n_{\text{in}}=25$ & $n_{\text{in}}=50$ & $n_{\text{in}}=100$ & $n_{\text{in}}=25$ & $n_{\text{in}}=50$ & $n_{\text{in}}=100$ \\
    \midrule
    \multicolumn{7}{@{}c}{\textit{Panel A: Single explanatory variable ($s=1$)}} \\
    \midrule
    INLS  & 0.1987 & 0.1441 & 0.1212 & 0.3858 & 0.3467 & 0.3292 \\
    CC-INLS (vertices)    & 0.1289 & 0.1217 & 0.1131 & 0.3536 & 0.3348 & 0.3264 \\
    CC-INLS (centroids) & 0.1271 & 0.1178 & 0.1107 & 0.3442 & 0.3266 & 0.3186 \\
    \midrule
    \multicolumn{7}{@{}c}{\textit{Panel B: Two explanatory variables ($s=2$)}} \\
    \midrule
    INLS  & 0.4464 & 0.3425 & 0.2565 & 0.5574 & 0.4773 & 0.4259 \\
    CC-INLS (vertices)    & 0.1782 & 0.1522 & 0.1504 & 0.3992 & 0.3952 & 0.3960 \\
    CC-INLS (centroids) & 0.1790 & 0.1509 & 0.1458 & 0.3931 & 0.3789 & 0.3715 \\
    \bottomrule
  \end{tabular}
  
  \vspace{1ex}
  \footnotesize{\textit{Note:} $n_{\text{in}}$ denotes the in-sample size used for model estimation and $n_{\text{out}}$ denotes the test set size.}
\end{table}

\begin{figure}[htp]
	\centering
	\begin{subfigure}[b]{0.49\textwidth}
		\centering
		\includegraphics[width=1\textwidth]{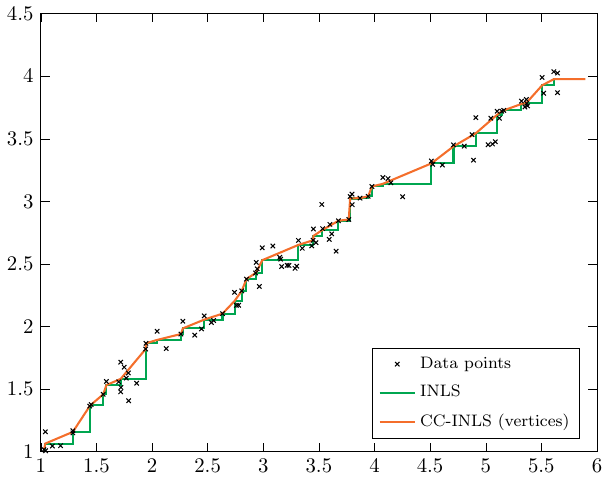} 
		\caption[]%
		{{\small Low noise (vertices)}}    
		\label{fig:CD_a}
	\end{subfigure}
	\begin{subfigure}[b]{0.49\textwidth}  
		\centering 
		\includegraphics[width=1\textwidth]{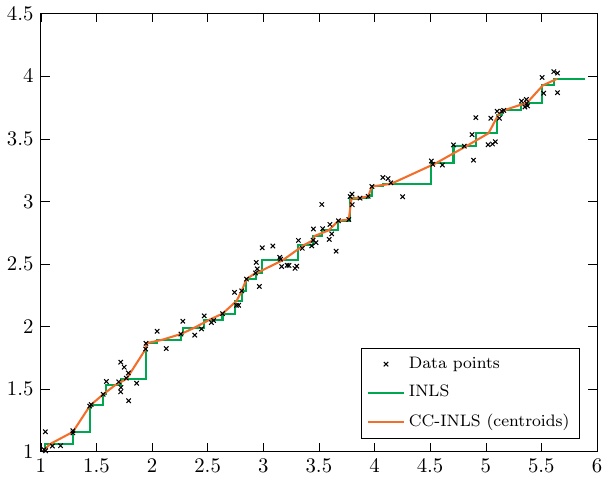} 
		\caption[]%
		{{\small Low noise (centroids)}}    
		\label{fig:CD_b}
	\end{subfigure}
	\begin{subfigure}[b]{0.49\textwidth}   
		\centering 
		\includegraphics[width=1\textwidth]{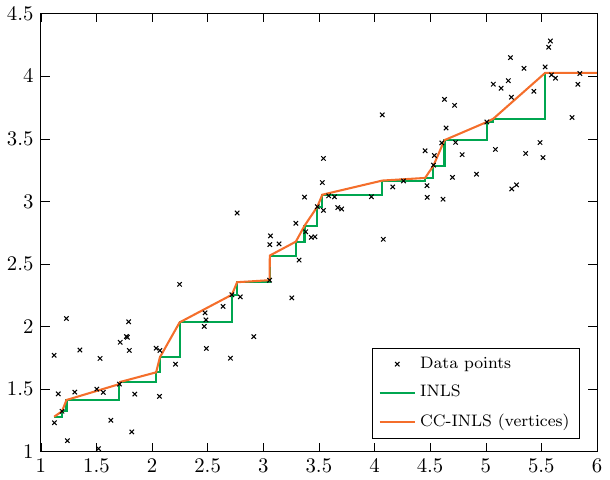} 
		\caption[]%
		{{\small High noise (vertices)}}    
		\label{fig:CD_c}
	\end{subfigure}
	\begin{subfigure}[b]{0.49\textwidth}   
		\centering 
		\includegraphics[width=1\textwidth]{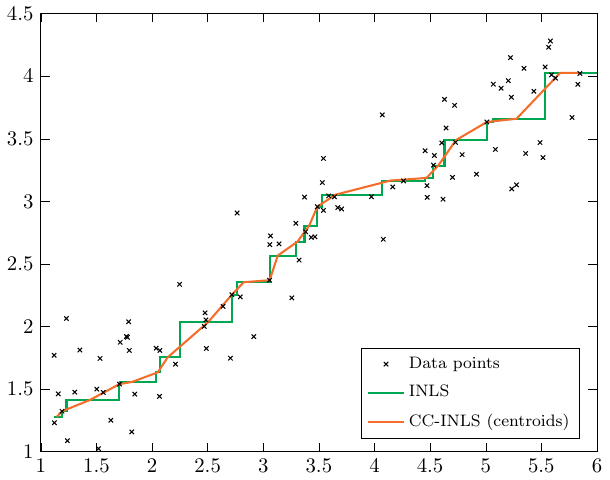} 
		\caption[]%
		{{\small High noise (centroids)}}    
		\label{fig:CD_d}
	\end{subfigure}
	\caption[]%
	{\centering Illustration of the Cobb-Douglas functions estimated by INLS, CC-INLS (vertices), CC-INLS (centroids) across low and high noise scenarios.}
	\label{fig:CD}
\end{figure}

Figure \ref{fig:CD} displays the fitted functions for a sample Cobb-Douglas DGP. The figure illustrates how both variants of the CC-INLS estimator (orange line) construct a continuous surface that smooths the discontinuous INLS step function (green line) while preserving the underlying monotonic pattern and diminishing marginal returns across both low and high noise scenarios.

\subsection{Summary of simulation results}

\begin{figure}[H]
    \centering
    \begin{subfigure}[b]{0.9\textwidth}
        \centering
        \includegraphics[width=0.95\textwidth]{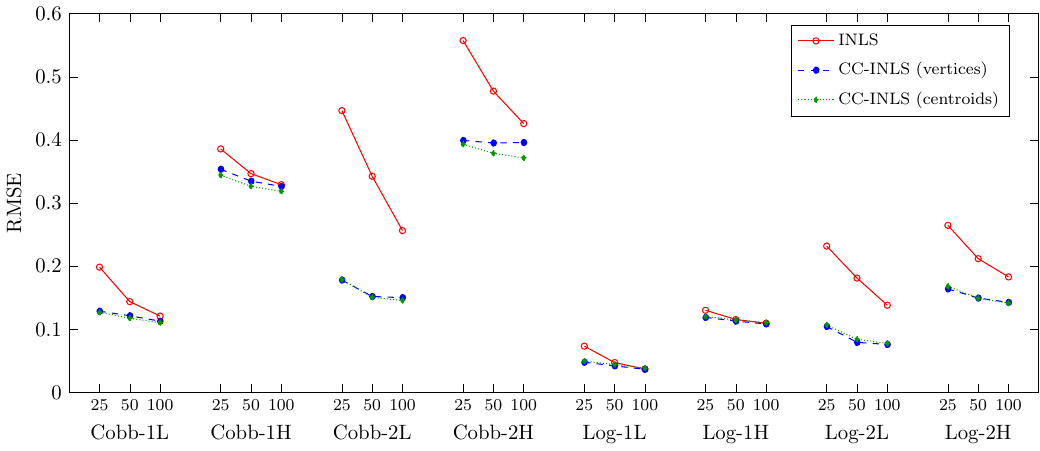}
        \caption{}
        \label{fig:rmse_comparison}
    \end{subfigure}
    
    \vspace{1em}
    
    \begin{subfigure}[b]{0.9\textwidth}
        \centering
        \hspace*{-0.3cm}\includegraphics[width=0.95\textwidth]{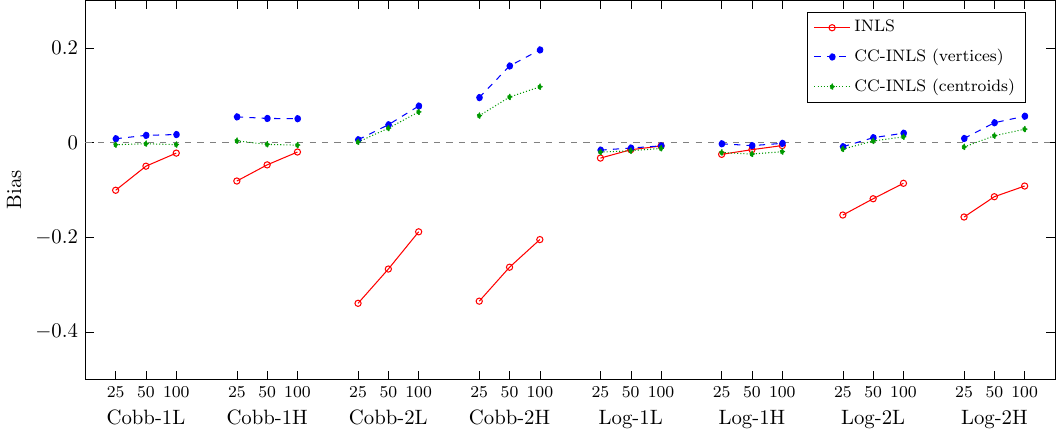}
        \caption{} 
        \label{fig:bias_comparison}
    \end{subfigure}
    \caption{Comparison of out-of-sample performance across all scenarios.}
    \label{fig:performance_comparison}
    
    \vspace{1ex}
    \footnotesize{\textit{Note:} The numbers 25, 50, and 100 on the horizontal axis represent the in-sample sizes used for model estimation. The abbreviations below them denote the simulated scenarios: the functional form (Cobb for Cobb-Douglas, Log for logistic), the dimensionality of the explanatory variables (1 for $s=1$, 2 for $s=2$), and the noise level (L for low noise, H for high noise).}
\end{figure}

To synthesize the finite-sample performance across all experimental settings, Figure \ref{fig:performance_comparison} visualizes the out-of-sample RMSE and bias.\footnote{Detailed MSE and bias statistics are deferred to Appendix \ref{sec:supplement_tables}.} The graphical summary confirms that both variants of the CC-INLS estimator systematically outperform the standard INLS step function in terms of RMSE across virtually all scenarios. The performance gap is particularly pronounced in multivariate settings ($s = 2$) and high-noise environments. Furthermore, the bias panel reveals a critical structural advantage: while the unsmoothed INLS estimator consistently exhibits a strong negative bias, the proposed piece-wise linear smoothing techniques effectively mitigate this downward bias, shifting the estimates significantly closer to zero.

Interestingly, while the Cobb-Douglas function represents a straightforward globally concave surface compared to the complex S-shape of the logistic curve, its overall estimation errors are generally higher. Despite this higher baseline error, the benefit of the proposed estimator remains robust, with the most pronounced improvements frequently occurring in the smallest training samples ($n_{\text{in}}=25$). This reaffirms that the conditional convexity framework is adaptable across fundamentally different underlying functional forms. 

\section{Application to Finnish municipalities}\label{sec:appl}

To illustrate the practical utility of the proposed CC-INLS estimator, we apply it to urban economics, specifically focusing on economies of agglomeration. Similar to firm-level economies of scale, the costs and benefits of agglomerating economic activity are hypothesized to increase as the agglomerated urban cluster becomes larger. Following a similar line of inquiry as \textcite{arauzo-carod_determinants_2007}, we utilize municipality-level panel data from Finland to estimate the functional relationship between the number of jobs (explanatory variable) and the total population of the municipality (dependent variable).\footnote{The data used in this application are publicly available from Statistics Finland at:  \url{https://pxdata.stat.fi/PxWeb/pxweb/en/StatFin/}.}

Our primary objectives in this empirical application are to (1) visually illustrate the shape of the estimated CC-INLS regression curves and (2) assess the out-of-sample performance of the two alternative variants of CC-INLS (vertices versus centroids).

To conduct a rigorous out-of-sample evaluation, we split the municipality-level panel data into a training set and a test set. The training set is constructed using the average population and job figures over the 2012--2016 period. To ensure the estimations reflect typical urban dynamics and avoid skewed boundaries, we exclude the capital city, Helsinki, as an extreme outlier, along with small municipalities with fewer than 10,000 inhabitants. This filtering process results in a training sample of 102 municipalities. For the out-of-sample evaluation, we utilize the annual data for these same 102 municipalities over the subsequent 2017--2021 period, yielding a test set of 510 distinct observations.

\begin{figure}[H]
    \centering
    \includegraphics[width=0.75\linewidth]{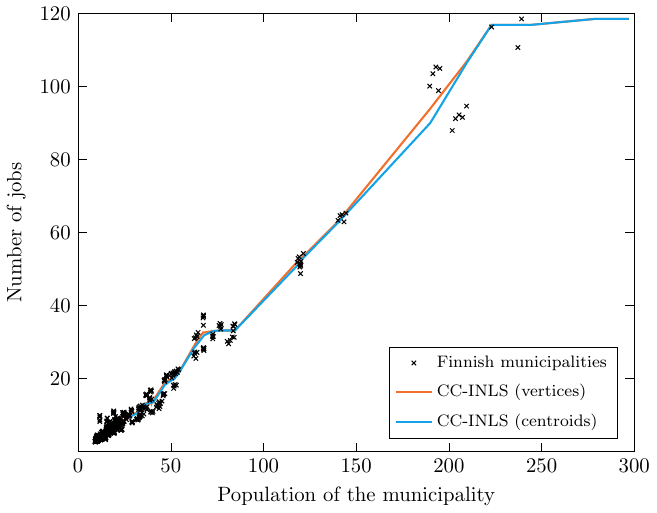}
    \caption{Illustration of the estimated CC-INLS regression functions in the test set.}
    \label{fig:CC-INLS}
\end{figure}

Figure \ref{fig:CC-INLS} plots the out-of-sample observations against the predicted regression surfaces. These continuous, piece-wise linear curves are constructed by applying two variants of the CC-INLS estimator, fitted on the 2012--2016 training data, to the population levels of the test set. Both CC-INLS variants construct out-of-sample predictions that closely align with the observed data trajectory. Consistent with the Monte Carlo simulation results, the empirical differences in predictive performance between using vertices and centroids as anchor points are negligible, yielding out-of-sample RMSE of 2.5191 and 2.5651, respectively.

A key advantage of our proposed methodology in this context is its ability to estimate marginal products and scale elasticities in a nonparametric fashion. The smoothed, piece-wise linear function provides a direct and meaningful estimate of the marginal increase in jobs associated with an increase in population, a value that the INLS step function cannot provide.

\section{Conclusions}\label{sec:conc}

This paper addresses a fundamental limitation of the standard INLS estimator: its discontinuous step-function nature, which precludes the evaluation of critical marginal economic properties. By integrating the concept of conditional convexity through a bilevel programming framework, we introduce the CC-INLS estimator. This approach successfully constructs a continuous, piece-wise linear regression surface that strictly enforces monotonicity while preserving localized shape variations and non-convexities identified in the data.

Our Monte Carlo simulations demonstrate that the proposed piece-wise linear smoothing technique does more than merely bridge the gap for marginal evaluations; it systematically enhances overall estimation accuracy. Across both non-convex (logistic) and strictly concave (Cobb-Douglas) DGPs, the CC-INLS estimator consistently yielded lower out-of-sample RMSE statistics and effectively mitigated the downward bias inherent in the unsmoothed INLS step function. These performance gains were particularly pronounced in multivariate settings and high-noise environments. 

Furthermore, the empirical application to agglomeration economies in Finnish municipalities corroborates these findings. The CC-INLS estimator not only provided highly accurate out-of-sample predictions but also demonstrated its practical value by enabling the direct, nonparametric estimation of scale elasticities. 

Notably, both the simulated and empirical results indicate that the choice of anchor points (vertices versus centroids) has a negligible impact on predictive performance, underscoring the robustness of the conditional convexity principle itself.

Several promising avenues remain for future research. First, while our single-level MILP reformulation allows for exact computation using standard solvers, the computational burden naturally scales with sample size and dimensionality. Developing tailored heuristic algorithms or efficient decomposition techniques to handle massive datasets represents an important practical extension. Second, establishing the asymptotic statistical properties, such as consistency and convergence rates, of the CC-INLS estimator would provide deeper theoretical foundations. Finally, adapting the bilevel conditional convexity framework to accommodate other complex shape constraints could further broaden its applicability across operational research and empirical economics.

\section*{Acknowledgments}\label{sec:ack}
This research was partially supported by the CIPROM/2024/34 grant, funded by the Conselleria de Educación, Cultura, Universidades y Empleo, Generalitat Valenciana.


\printbibliography

@article{wang_nonparametric_2026,
	title = {A nonparametric least-squares model in network data envelopment analysis},
	volume = {331},
	issn = {0377-2217},
	url = {https://www.sciencedirect.com/science/article/pii/S0377221725007994},
	doi = {10.1016/j.ejor.2025.09.045},
	number = {2},
	urldate = {2026-05-13},
	journal = {European Journal of Operational Research},
	author = {Wang, Zixuan and Yang, Min and Liang, Liang and Zhu, Joe},
	month = jun,
	year = {2026},
	pages = {555--569},
}

@article{luo_sparse_2026,
	title = {Sparse convex quantile regression: {A} generalized benders decomposition approach},
	issn = {0377-2217},
	shorttitle = {Sparse convex quantile regression},
	url = {https://www.sciencedirect.com/science/article/pii/S0377221726002092},
	doi = {10.1016/j.ejor.2026.03.001},
	urldate = {2026-05-13},
	journal = {European Journal of Operational Research},
	author = {Luo, Xiaoyu and Gao, Chuanhou},
	month = mar,
	year = {2026},
}

@article{dai_optimal_2025,
	title = {Optimal resource allocation: {Convex} quantile regression approach},
	volume = {324},
	issn = {0377-2217},
	shorttitle = {Optimal resource allocation},
	url = {https://www.sciencedirect.com/science/article/pii/S0377221725000037},
	doi = {10.1016/j.ejor.2025.01.003},
	number = {1},
	urldate = {2026-05-13},
	journal = {European Journal of Operational Research},
	author = {Dai, Sheng and Kuosmanen, Natalia and Kuosmanen, Timo and Liesiö, Juuso},
	month = jul,
	year = {2025},
	pages = {221--230},
}

@article{keshvari_stochastic_2013,
	title = {Stochastic non-convex envelopment of data: {Applying} isotonic regression to frontier estimation},
	volume = {231},
	issn = {03772217},
	shorttitle = {Stochastic non-convex envelopment of data},
	url = {https://linkinghub.elsevier.com/retrieve/pii/S0377221713004748},
	doi = {10.1016/j.ejor.2013.06.005},
	language = {en},
	number = {2},
	urldate = {2025-12-31},
	journal = {European Journal of Operational Research},
	author = {Keshvari, Abolfazl and Kuosmanen, Timo},
	month = dec,
	year = {2013},
	pages = {481--491},
}

@article{fang_projection_2021,
	title = {A {Projection} {Framework} for {Testing} {Shape} {Restrictions} {That} {Form} {Convex} {Cones}},
	volume = {89},
	copyright = {© 2021 The Econometric Society},
	issn = {1468-0262},
	url = {https://onlinelibrary.wiley.com/doi/abs/10.3982/ECTA17764},
	doi = {10.3982/ECTA17764},
	language = {en},
	number = {5},
	urldate = {2025-11-15},
	journal = {Econometrica},
	author = {Fang, Zheng and Seo, Juwon},
	year = {2021},
	note = {\_eprint: https://onlinelibrary.wiley.com/doi/pdf/10.3982/ECTA17764},
	pages = {2439--2458},
}

@article{mukherjee_least_2024,
	title = {Least squares estimation of a quasiconvex regression function},
	volume = {86},
	issn = {1369-7412},
	url = {https://doi.org/10.1093/jrsssb/qkad133},
	doi = {10.1093/jrsssb/qkad133},
	number = {2},
	urldate = {2025-11-15},
	journal = {Journal of the Royal Statistical Society Series B: Statistical Methodology},
	author = {Mukherjee, Somabha and Patra, Rohit K and Johnson, Andrew L and Morita, Hiroshi},
	month = apr,
	year = {2024},
	pages = {512--534},
}

@article{rodseth_combining_2025,
	title = {Combining convex regression with the regression discontinuity design: {Effectiveness} of e-scooter providers during the {Covid}-19 lockdown},
	issn = {0377-2217},
	shorttitle = {Combining convex regression with the regression discontinuity design},
	url = {https://www.sciencedirect.com/science/article/pii/S0377221725009051},
	doi = {10.1016/j.ejor.2025.11.011},
	urldate = {2025-11-15},
	journal = {European Journal of Operational Research},
	author = {Rødseth, Kenneth Løvold and Kuosmanen, Timo and Holmen, Rasmus Bøgh},
	month = nov,
	year = {2025},
}

@article{monge_measuring_2025,
	title = {Measuring {Efficiency} in {Data} {Envelopment} {Analysis} {Under} {Conditional} {Convexity}},
	volume = {205},
	issn = {1573-2878},
	url = {https://doi.org/10.1007/s10957-025-02659-8},
	doi = {10.1007/s10957-025-02659-8},
	language = {en},
	number = {3},
	urldate = {2025-11-15},
	journal = {Journal of Optimization Theory and Applications},
	author = {Monge, Juan F. and Ruiz, José L.},
	month = apr,
	year = {2025},
	pages = {46},
}

@article{monge_setting_2023,
	title = {Setting closer targets based on non-dominated convex combinations of {Pareto}-efficient units: {A} bi-level linear programming approach in {Data} {Envelopment} {Analysis}},
	volume = {311},
	issn = {0377-2217},
	shorttitle = {Setting closer targets based on non-dominated convex combinations of {Pareto}-efficient units},
	url = {https://www.sciencedirect.com/science/article/pii/S0377221723004198},
	doi = {10.1016/j.ejor.2023.05.034},
	number = {3},
	urldate = {2025-11-15},
	journal = {European Journal of Operational Research},
	author = {Monge, Juan F. and Ruiz, José L.},
	month = dec,
	year = {2023},
	pages = {1084--1096},
}

@article{feng_nonparametric_2022,
	title = {Nonparametric, {Tuning}-{Free} {Estimation} of {S}-{Shaped} {Functions}},
	volume = {84},
	issn = {1369-7412},
	url = {https://doi.org/10.1111/rssb.12481},
	doi = {10.1111/rssb.12481},
	number = {4},
	urldate = {2025-11-15},
	journal = {Journal of the Royal Statistical Society Series B: Statistical Methodology},
	author = {Feng, Oliver Y. and Chen, Yining and Han, Qiyang and Carroll, Raymond J. and Samworth, Richard J.},
	month = sep,
	year = {2022},
	pages = {1324--1352},
}

@article{kuosmanen_dea_2001,
	series = {Data {Envelopment} {Analysis}},
	title = {{DEA} with efficiency classification preserving conditional convexity},
	volume = {132},
	issn = {0377-2217},
	url = {https://www.sciencedirect.com/science/article/pii/S0377221700001557},
	doi = {10.1016/S0377-2217(00)00155-7},
	number = {2},
	urldate = {2024-12-20},
	journal = {European Journal of Operational Research},
	author = {Kuosmanen, Timo},
	month = jul,
	year = {2001},
	pages = {326--342},
}

@article{curmei_shape-constrained_2025,
	title = {Shape-{Constrained} {Regression} {Using} {Sum} of {Squares} {Polynomials}},
	volume = {73},
	issn = {0030-364X},
	url = {https://pubsonline.informs.org/doi/abs/10.1287/opre.2021.0383},
	doi = {10.1287/opre.2021.0383},
	number = {1},
	urldate = {2025-11-11},
	journal = {Operations Research},
	publisher = {INFORMS},
	author = {Curmei, Mihaela and Hall, Georgina},
	month = jan,
	year = {2025},
	pages = {543--559},
}

@article{arauzo-carod_determinants_2007,
	title = {Determinants of population and jobs at a local level},
	volume = {41},
	issn = {1432-0592},
	url = {https://doi.org/10.1007/s00168-006-0058-6},
	doi = {10.1007/s00168-006-0058-6},
	language = {en},
	number = {1},
	urldate = {2025-11-11},
	journal = {The Annals of Regional Science},
	author = {Arauzo-Carod, Josep-Maria},
	month = mar,
	year = {2007},
	pages = {87--104},
}

@article{hildreth_point_1954,
	title = {Point {Estimates} of {Ordinates} of {Concave} {Functions}},
	volume = {49},
	issn = {0162-1459},
	url = {https://www.jstor.org/stable/2281132},
	doi = {10.2307/2281132},
	number = {267},
	urldate = {2025-11-11},
	journal = {Journal of the American Statistical Association},
	publisher = {[American Statistical Association, Taylor \& Francis, Ltd.]},
	author = {Hildreth, Clifford},
	year = {1954},
	pages = {598--619},
}

@article{brunk_maximum_1955,
	title = {Maximum {Likelihood} {Estimates} of {Monotone} {Parameters}},
	volume = {26},
	issn = {0003-4851},
	url = {https://www.jstor.org/stable/2236374},
	number = {4},
	urldate = {2025-11-11},
	journal = {The Annals of Mathematical Statistics},
	publisher = {Institute of Mathematical Statistics},
	author = {Brunk, H. D.},
	year = {1955},
	pages = {607--616},
}

@article{dai_can_2025,
	title = {Can {Omitted} {Carbon} {Abatement} {Explain} {Productivity} {Stagnation}?},
	volume = {71},
	copyright = {Creative Commons Attribution 4.0 International Licence},
	issn = {0034-6586},
	url = {https://onlinelibrary.wiley.com/doi/10.1111/roiw.70012},
	doi = {10.1111/roiw.70012},
	number = {2},
	urldate = {2025-03-20},
	journal = {Review of Income and Wealth},
	publisher = {John Wiley \& Sons, Ltd},
	author = {Dai, Sheng and Kuosmanen, Timo and Zhou, Xun},
	month = may,
	year = {2025},
	pages = {e70012},
}

@article{Kuosmanen2008,
	title = {Representation theorem for convex nonparametric least squares},
	volume = {11},
	issn = {13684221},
	url = {http://doi.wiley.com/10.1111/j.1368-423X.2008.00239.x},
	doi = {10.1111/j.1368-423X.2008.00239.x},
	number = {2},
	urldate = {2018-05-27},
	journal = {Econometrics Journal},
	publisher = {Wiley/Blackwell (10.1111)},
	author = {Kuosmanen, Timo},
	month = jul,
	year = {2008},
	note = {ISBN: 1368-423X},
	pages = {308--325},
}

@article{Kuosmanen2021,
	title = {Shadow prices and marginal abatement costs: {Convex} quantile regression approach},
	volume = {289},
	copyright = {All rights reserved},
	issn = {03772217},
	doi = {10.1016/j.ejor.2020.07.036},
	number = {2},
	urldate = {2020-11-04},
	journal = {European Journal of Operational Research},
	publisher = {Elsevier B.V.},
	author = {Kuosmanen, Timo and Zhou, Xun},
	month = mar,
	year = {2021},
	pages = {666--675},
}
\baselineskip 12pt


\newpage
\begin{appendices}
\section*{Appendix}\label{sec:app}
\baselineskip 20pt
\renewcommand{\thesubsection}{\Alph{subsection}}
\setcounter{table}{0}
\setcounter{figure}{0}
\setcounter{equation}{0}
\setcounter{theorem}{0}
\renewcommand{\theequation}{A\arabic{equation}} 
\renewcommand{\thetable}{A\arabic{table}} 
\renewcommand{\thefigure}{A\arabic{figure}} 

\subsection{Supplementary tables}
\label{sec:supplement_tables}

\begin{table}[H]
  \centering
  \caption{Out-of-sample MSE for the logistic DGPs ($n_{\text{out}}=100$).}
  \label{tab:mse_logistic}
  \renewcommand{\arraystretch}{1.2}
  \small
  \begin{tabular}{@{}lcccccc@{}}
    \toprule
    \multirow{2}{*}{Estimator} & \multicolumn{3}{c}{Low noise ($\sigma_\varepsilon = 0.03$)} & \multicolumn{3}{c}{High noise ($\sigma_\varepsilon = 0.1$)} \\
    \cmidrule(lr){2-4} \cmidrule(lr){5-7}
    & $n_{\text{in}}=25$ & $n_{\text{in}}=50$ & $n_{\text{in}}=100$ & $n_{\text{in}}=25$ & $n_{\text{in}}=50$ & $n_{\text{in}}=100$ \\
    \midrule
    \multicolumn{7}{@{}c}{\textit{Panel A: Single explanatory variable ($s=1$)}} \\
    \midrule
    INLS  & 0.0060 & 0.0023 & 0.0014 & 0.0174 & 0.0135 & 0.0121 \\
    CC-INLS (vertices)    & 0.0024 & 0.0018 & 0.0014 & 0.0143 & 0.0129 & 0.0119 \\
    CC-INLS (centroids) & 0.0025 & 0.0020 & 0.0015 & 0.0147 & 0.0133 & 0.0123 \\
    \midrule
    \multicolumn{7}{@{}c}{\textit{Panel B: Two explanatory variables ($s=2$)}} \\
    \midrule
    INLS  & 0.0561 & 0.0340 & 0.0198 & 0.0730 & 0.0465 & 0.0344 \\
    CC-INLS (vertices)    & 0.0129 & 0.0074 & 0.0066 & 0.0284 & 0.0230 & 0.0209 \\
    CC-INLS (centroids) & 0.0135 & 0.0084 & 0.0069 & 0.0300 & 0.0231 & 0.0207 \\
    \bottomrule
  \end{tabular}
  
  \vspace{1ex}
  \footnotesize{\textit{Note:} $n_{\text{in}}$ denotes the in-sample size used for model estimation and $n_{\text{out}}$ denotes the test set size.}
\end{table}

\begin{table}[H]
  \centering
  \caption{Out-of-sample bias for the logistic DGPs ($n_{\text{out}}=100$).}
  \label{tab:bias_logistic}
  \renewcommand{\arraystretch}{1.2}
  \small
  \makebox[\textwidth][c]{
  \begin{tabular}{@{}lcccccc@{}}
    \toprule
    \multirow{2}{*}{Estimator} & \multicolumn{3}{c}{Low noise ($\sigma_\varepsilon = 0.03$)} & \multicolumn{3}{c}{High noise ($\sigma_\varepsilon = 0.1$)} \\
    \cmidrule(lr){2-4} \cmidrule(lr){5-7}
    & $n_{\text{in}}=25$ & $n_{\text{in}}=50$ & $n_{\text{in}}=100$ & $n_{\text{in}}=25$ & $n_{\text{in}}=50$ & $n_{\text{in}}=100$ \\
    \midrule
    \multicolumn{7}{@{}c}{\textit{Panel A: Single explanatory variable ($s=1$)}} \\
    \midrule
    INLS  & -0.0328 & -0.0152 & -0.0073 & -0.0249 & -0.0150 & -0.0064 \\
    CC-INLS (vertices)    & -0.0160 & -0.0116 & -0.0070 & -0.0026 & -0.0065 & -0.0017 \\
    CC-INLS (centroids) & -0.0201 & -0.0174 & -0.0124 & -0.0211 & -0.0243 & -0.0193 \\
    \midrule
    \multicolumn{7}{@{}c}{\textit{Panel B: Two explanatory variables ($s=2$)}} \\
    \midrule
    INLS  & -0.1530 & -0.1186 & -0.0860 & -0.1572 & -0.1143 & -0.0918 \\
    CC-INLS (vertices)    & -0.0088 & 0.0102 & 0.0197 & 0.0086 & 0.0418 & 0.0556 \\
    CC-INLS (centroids) & -0.0138 & 0.0032 & 0.0121 & -0.0094 & 0.0143 & 0.0282 \\
    \bottomrule
  \end{tabular}
  }
  
  \vspace{1ex}
  \footnotesize{\textit{Note:} $n_{\text{in}}$ denotes the in-sample size used for model estimation and $n_{\text{out}}$ denotes the test set size.}
\end{table}

\begin{table}[H]
  \centering
  \caption{Out-of-sample MSE for the Cobb-Douglas DGPs ($n_{\text{out}}=100$).}
  \label{tab:mse_cd}
  \renewcommand{\arraystretch}{1.2}
  \small
  \begin{tabular}{@{}lcccccc@{}}
    \toprule
    \multirow{2}{*}{Estimator} & \multicolumn{3}{c}{Low noise ($\sigma_\varepsilon = 0.1$)} & \multicolumn{3}{c}{High noise ($\sigma_\varepsilon = 0.2$)} \\
    \cmidrule(lr){2-4} \cmidrule(lr){5-7}
    & $n_{\text{in}}=25$ & $n_{\text{in}}=50$ & $n_{\text{in}}=100$ & $n_{\text{in}}=25$ & $n_{\text{in}}=50$ & $n_{\text{in}}=100$ \\
    \midrule
    \multicolumn{7}{@{}c}{\textit{Panel A: Single explanatory variable ($s=1$)}} \\
    \midrule
    INLS  & 0.0407 & 0.0211 & 0.0148 & 0.1511 & 0.1208 & 0.1090 \\
    CC-INLS (vertices)    & 0.0170 & 0.0149 & 0.0129 & 0.1265 & 0.1127 & 0.1070 \\
    CC-INLS (centroids) & 0.0165 & 0.0140 & 0.0123 & 0.1197 & 0.1071 & 0.1020 \\
    \midrule
    \multicolumn{7}{@{}c}{\textit{Panel B: Two explanatory variables ($s=2$)}} \\
    \midrule
    INLS  & 0.2035 & 0.1189 & 0.0665 & 0.3184 & 0.2314 & 0.1837 \\
    CC-INLS (vertices)    & 0.0331 & 0.0240 & 0.0230 & 0.1617 & 0.1587 & 0.1583 \\
    CC-INLS (centroids) & 0.0335 & 0.0236 & 0.0217 & 0.1565 & 0.1457 & 0.1394 \\
    \bottomrule
  \end{tabular}
  
  \vspace{1ex}
  \footnotesize{\textit{Note:} $n_{\text{in}}$ denotes the in-sample size used for model estimation and $n_{\text{out}}$ denotes the test set size.}
\end{table}

\begin{table}[H]
  \centering
  \caption{Out-of-sample bias for the Cobb-Douglas DGPs ($n_{\text{out}}=100$).}
  \label{tab:bias_cd}
  \renewcommand{\arraystretch}{1.2}
  \small
  \makebox[\textwidth][c]{
  \begin{tabular}{@{}lcccccc@{}}
    \toprule
    \multirow{2}{*}{Estimator} & \multicolumn{3}{c}{Low noise ($\sigma_\varepsilon = 0.1$)} & \multicolumn{3}{c}{High noise ($\sigma_\varepsilon = 0.2$)} \\
    \cmidrule(lr){2-4} \cmidrule(lr){5-7}
    & $n_{\text{in}}=25$ & $n_{\text{in}}=50$ & $n_{\text{in}}=100$ & $n_{\text{in}}=25$ & $n_{\text{in}}=50$ & $n_{\text{in}}=100$ \\
    \midrule
    \multicolumn{7}{@{}c}{\textit{Panel A: Single explanatory variable ($s=1$)}} \\
    \midrule
    INLS  & -0.1007 & -0.0499 & -0.0223 & -0.0812 & -0.0471 & -0.0200 \\
    CC-INLS (vertices)    & 0.0080 & 0.0152 & 0.0169 & 0.0542 & 0.0509 & 0.0503 \\
    CC-INLS (centroids) & -0.0046 & -0.0025 & -0.0045 & 0.0036 & -0.0039 & -0.0055 \\
    \midrule
    \multicolumn{7}{@{}c}{\textit{Panel B: Two explanatory variables ($s=2$)}} \\
    \midrule
    INLS  & -0.3393 & -0.2670 & -0.1884 & -0.3348 & -0.2631 & -0.2048 \\
    CC-INLS (vertices)    & 0.0059 & 0.0375 & 0.0770 & 0.0949 & 0.1616 & 0.1955 \\
    CC-INLS (centroids) & 0.0012 & 0.0302 & 0.0646 & 0.0566 & 0.0960 & 0.1176 \\
    \bottomrule
  \end{tabular}
  }
  
  \vspace{1ex}
  \footnotesize{\textit{Note:} $n_{\text{in}}$ denotes the in-sample size used for model estimation and $n_{\text{out}}$ denotes the test set size.}
\end{table}


\end{appendices}

\end{document}